\documentclass[runningheads]{svmult}

\usepackage{makeidx}   
\usepackage{graphicx}  
\usepackage{subeqnar}  
\usepackage{multicol}  
\usepackage{physprbb}  
\makeindex             

\begin{document}
\title*{The Extreme Ends of the Metallicity Distribution in dSph Galaxies}
\toctitle{The Extreme Ends of the Metallicity Distribution in dSph Galaxies}
%
%
\titlerunning{Metallicity of dSph Galaxies}
%
\author{Matthew Shetrone\inst{1}}
\authorrunning{Matthew Shetrone }
%
%
\institute{University of Texas, McDonald Observatory, Fort Davis, Tx 79734, USA}

\maketitle              

\begin{abstract}

This paper reviews recent abundance results of local dSph giants.  All dSph
systems seem to show evidence of slow star formation rates, compared to 
the Milky Way,  based on the most metal-rich stars exhibiting low [even-Z/Fe]
ratios and high [s-process/r-process] ratios.   The most metal-poor
stars in the Draco, Ursa Minor, Sextans and Sculptor dSphs seem to show a 
split in their light even-Z, Mg and O, and the heavier even-Z, Ca and Ti, 
abundance ratios, where the light even-Z are halo like and the heavier
even-Z elements exhibit sub-halo abundance ratios.   This split remains a
mystery.   A review of the first dSph abundance results from 
FLAMES$+$GIRAFFE on the VLT and 
HRS on the Hobby-Eberly Telescope \footnote{The Hobby-Eberly Telescope (HET) 
is a joint project of the University of Texas at Austin, the Pennsylvania 
State University, Stanford University, Ludwig-Maximilians-Universit\"at
M\"unchen, and Georg-August-Universit\"at G\"ottingen. 
The HET is named in honor of its principal benefactors, William P. Hobby 
and Robert E. Eberly.}
shows that the study of chemical evolution of dSph galaxies is rapidly 
moving out of infancy and into an era requiring very large surveys and/or 
targeted studies.

\end{abstract}

\section{The Most Metal-rich dSph Stars}

The most metal-rich stars in dwarf spheroidals (dSph) have been shown to have 
significantly lower even-Z abundance ratios than stars of similar metallicity 
in the Milky Way (MW).  In addition, the most metal-rich dSph stars are 
dominated by an s-process abundance pattern in comparison to stars of 
similar metallicity in the MW.
This has been interpreted as excessive contamination by Type Ia super-novae (SN)
and asymptotic giant branch (AGB) stars
( Bonifacio et al. 2000, Shetrone et al. 2001, Smecker-Hane \& McWilliam 2002).
By comparing these results to MW chemical evolution,
Lanfranchi \& Matteucci (2003) conclude that the dSph galaxies have had a 
slower star formation rate
than the MW (Lanfranchi \& Matteucci 2003).  This slow star formation, when
combined with an efficient galactic wind, allows 
the contribution of Type Ia SN and AGB stars to be incorporated into the ISM
before the Type II SN can bring the metallicity up to MW thick disk 
metallicities.

Recent abundance ratio work in this field falls into two 
categories.  The first category has been investigations into 
aspects of metal-poor AGB and Type Ia SN yields and their relationship to 
the chemical evolution in the dSph galaxies, e.g. McWilliam et al. (2003), 
Venn et al. (2004), McWilliam \& Smecker-Hane (2005).   In these works
the abundances of specific elements are compared to predictions of  
yields of low metallicity SN and AGB stars.   While the origins of these
elements, such as Mn, Cu
and Y, may seem slightly esoteric, these types of analyses will help constrain 
future models of SN yields.   

The second catagory of dSph abundance investigations has been attempts to 
gain large enough samples to accurately 
model the extent of the chemical evolution, the relative contributions
the Type Ia and AGB yields and to what extent galactic winds have played a 
roll in the chemical evolution.
The new instruments that have come on-line in the last year have increased
the multiplexing capabilities of these surveys.  As an example, Figure 1
shows preliminary results from the Dwarf Abundance and Radial velocity Team 
(DART) ESO large program;  using UVES FLAMES$+$GIRAFFE on a sample of Sculptor
dSph giants, Hill (private communication, 2005) collected nearly 100 stellar 
spectra,
a sample larger than all of the literature high resolution dSph surveys 
combined.  This survey will be able to show subtle declines and trends that the 
other surveys could never detect.   The decline seen in [Ca/Fe] with 
increasing [Fe/H] reported by Shetrone et al. (2003) and Geisler et al. (2004)
are easily detected.  The spread in [Ca/Fe] at a given metallicity is being
investigated by DART.

\begin{figure}[1]
\begin{center}
\includegraphics[width=0.81\textwidth]{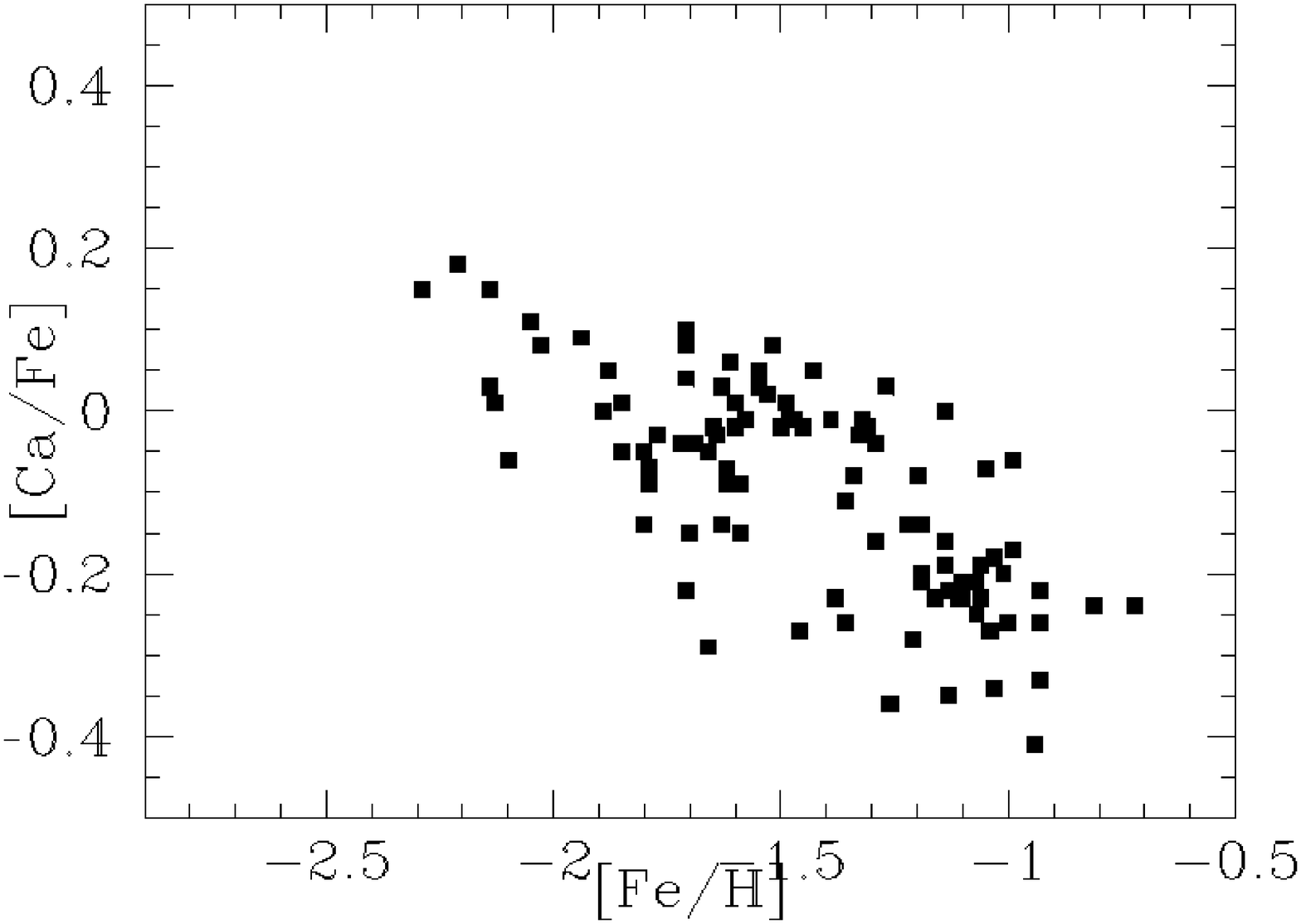}
\end{center}
\caption[]{An example of the new data sets coming available with the VLT$+$FLAMES.
  This data, from Hill (private communication, 2005) exhibits the steep 
decline of [Ca/Fe] with 
increasing [Fe/H] seen in all dSph.  In addition, the most metal-poor stars 
do not exhibit the typical MW halo ratio near 0.3 dex.}
\label{shetrone1}
\end{figure}

\section{The Most Metal-poor dSph Stars}

In a closed box or leaky-box chemical evolution model the most metal-poor 
stars would have formed before the majority of the Type Ia SN or the AGB
stars were significant contributors to the the ISM.  Thus, the abundances of
these very metal-poor stars should be excellent surrogates of the Pop III
and very metal-poor Pop II Type II SN products held in the dSph gravitational
potential.   This last point is important because if the yields of some
masses of Type II SN are lost from the dSph gravitational potential then 
this might significantly impact abundance ratios found in the next 
generation of dSph stars.

The first papers in this field suggested that the overall alpha abundances
found in the most metal-poor dSph stars are not similar to those found in the 
halo.  However, a more detailed analysis of the individual elements 
including corrections for differences in log gf values has shown that the 
O and Mg abundances are consistent with those found in the MW halo, while
the Ca and Ti abundances are systematically lower than those in the MW
halo, see figure 1, 2 and 6 in Shetrone (2004).    This can also be seen
in Figure 1 which shows that the most metal-poor Sculptor stars have 
[Ca/Fe] ratios less than
0.2 dex while the halo median at this metallicity is roughly 0.15 dex larger.

The difficulty with studies of the most metal-poor dSph stars is actually 
finding the most metal-poor stars.   Not only are these stars less numerous
than their more metal-rich counter parts, but their spatial distribution is
larger than the more metal-rich stars, e.g. Tolstoy et al. (2004), 
Palma et al. (2003), Harbeck et al. (2001), Majewski et al. (1999).   The
use of high resolution multi-object spectrographs such as FLAMES becomes 
less efficient in the search for the most-metal poor stars because of the 
large spatial extent, rarity and huge background contamination.   For these
types of studies targeting single star high resolution spectral 
follow-up to photometric or low resolution surveys can be more appropriate.

One star analyzed in the Shetrone et al. (1998) survey was found to be very 
metal-poor but with fairly low overall-alpha abundances.  Unfortunately, due to
the low S/N and low metallicity many of the elements had upper limits and large
error bars.  This single star, Draco 119, was re-observed by Fulbright
et al. (2004) with much higher S/N.  They were able to confirm that the 
Mg abundance was halo like while the Ca and Ti abundances were lower than
those found in the halo by a few tenths of dex.   Even more remarkable were
the upper limits found for the neutron capture elements.   Fulbright et al. 
found upper limits for [Ba/Fe] 1 dex lower than MW halo giants of similar 
metallicity,   and, even more amazingly, the upper limit for [Sr/Fe] was 
found to be nearly 2 dex lowerthan similiar MW giants.   
This begged the question: is the Draco 119 
abundance pattern unique, ie. due to some strange inhomogeneous mixing event,
or is this pattern found in all very metal-poor Draco stars.   

A search for equally metal-poor Draco stars, using the Hobby-Eberly telescope,
did not turn up any Draco giants as metal-poor as Draco 119; but it 
did turn up a few stars just
a few tenths more metal-rich.  By integrating long enough to detect the 
strong red Ba lines in these stars using the HRS on the Hobby-Eberly telescope,
Shetrone et al. (2005) did not find extremely 
neutron capture poor stars, see Figure 2.
The abundance pattern of Draco 119 appears to be a due to inhomogeneous mixing
and is not found in all very metal-poor Draco stars.

I would like to thank Kim Venn, Andy McWilliam, Verne Smith, Jon Fulbright
and the DART for preprints and invaluable discussions.  I would also like 
to thank the NSF for support through AST-0306884; and summer REU intern John 
Moore and the team at the HET for their assistance in bringing these results to
press.

\begin{figure}[2]
\begin{center}
\includegraphics[width=0.81\textwidth]{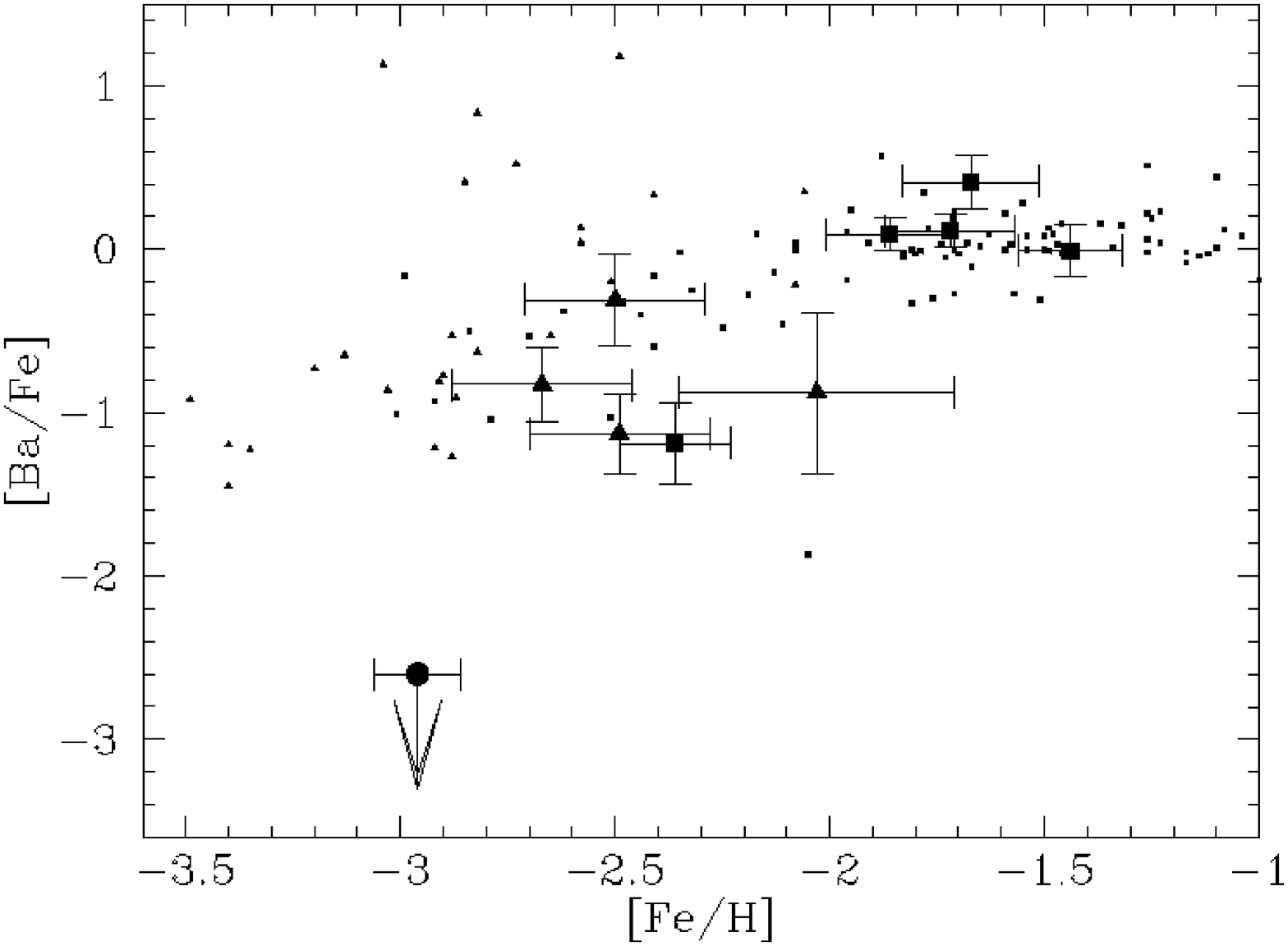}
\end{center}
\caption[]{The abundance ratio of Ba from the Shetrone et al. (1998), squares,
Fulbright et al. (2004), circle, and Shetrone et al. (2005), triangles shown
plotted against their derived metallicities.   The upper limit from 
Fulbright et al. (2004) is more than an order of magnitude lower than that
of the slightly more metal rich Draco dSph stars or comparable metallicity
MW halo stars, small symbols.}
\label{shetrone2}
\end{figure}

%

\end{document}